\documentclass{article}
\usepackage[T1]{fontenc} 
\usepackage[utf8]{inputenc} 
\usepackage{ismir,amsmath,cite,url}
\usepackage{graphicx}
\usepackage{color}
\usepackage{booktabs}

\usepackage{amsfonts}

\usepackage{lineno}

\title{Musika!\\Fast Infinite Waveform Music Generation}

\oneauthor
 {Marco Pasini \hspace{11em} Jan Schlüter}
 {Institute of Computational Perception, Johannes Kepler University Linz, Austria \\ {\tt marco.pasini.98@gmail.com \qquad jan.schlueter@jku.at}}

\def\authorname{M. Pasini, and J. Schlüter}

\usepackage[bookmarks=false,pdfauthor={\authorname},pdfsubject={\papersubject},hidelinks]{hyperref}

\begin{document}

\maketitle
\begin{abstract}
Fast and user-controllable music generation could enable novel ways of composing or performing music. However, state-of-the-art music generation systems require large amounts of data and computational resources for training, and are slow at inference. This makes them impractical for real-time interactive use. In this work, we introduce Musika, a music generation system that can be trained on hundreds of hours of music using a single consumer GPU, and that allows for much faster than real-time generation of music of arbitrary length on a consumer CPU. We achieve this by first learning a compact invertible representation of spectrogram magnitudes and phases with adversarial autoencoders, then training a Generative Adversarial Network (GAN) on this representation for a particular music domain. A latent coordinate system enables generating arbitrarily long sequences of excerpts in parallel, while a global context vector allows the music to remain stylistically coherent through time. We perform quantitative evaluations to assess the quality of the generated samples and showcase options for user control in piano and techno music generation. We release the source code and pretrained autoencoder weights at \href{https://github.com/marcoppasini/musika}{github.com/marcoppasini/musika}, such that a GAN can be trained on a new music domain with a single GPU in a matter of hours.
\end{abstract}
\section{Introduction}\label{sec:introduction}

Generating raw audio remains a difficult task to perform, considering the high temporal dimensionality of waveforms. Recently, a number of techniques based on deep learning architectures have been proposed: however, they often present limitations such as low generated music quality, lack of general coherence between distant time frames and slow generation speed. Regarding unconditional audio generation, autoregressive models are able to generate high quality audio with long-range dependencies; however, the sampling process is extremely slow and inefficient, which hinders possible real-world applications. On the other hand, non-autoregressive models can reach real-time generation, while they struggle to synthesize samples with satisfactory sound quality and are only able to generate samples of a fixed duration.

Considering the current shortcomings of non-autoregressive audio generation systems, in this work we propose Musika, a GAN-based system that allows fast unconditional and conditional generation of audio of arbitrary length. We achieve this by combining the following contributions: 

\begin{itemize}
    \item The use of a raw audio autoencoder which allows to encode samples into lower-dimensional invertible representations that are easier to model. We engineer the autoencoder with the specific goal of maximizing inference speed and minimizing training time, by relying on the generation of magnitude and phase spectrograms with low temporal resolution and an efficient adversarial training process
    \item The use of a latent coordinate system for the task of infinite-length audio generation
    \item The addition of a global style conditioning which allows the infinite-length generated samples to be stylistically coherent through time
    \item The possibility to perform both unconditional and conditional generation, with a variety of different conditioning signals, such as note density and tempo information
\end{itemize}
By avoiding auto-regression, generation can be fully parallelized and works much faster than real-time even on CPU.

\section{Related Work}\label{sec:related_work}

A popular family of generative models for audio consists in autoregressive models, such as WaveNet \cite{wavenet}, SampleRNN \cite{samplernn} and Jukebox \cite{jukebox}. WaveNet was the first model to show that autoregressive generation of raw audio waveforms is possible, and uses dilated convolutions to acquire a large receptive field over the input sample. SampleRNN is a model consisting of a hierarchical stack of recurrent units that are able to model the waveform at different resolutions, and thus capture a larger context and reduce the computational cost required to model the next sample. Jukebox uses a hierarchical VQVAE \cite{vqvae,vqvae2} to encode raw samples into a sequence of discrete codes at different levels. It then uses autoregressive transformers \cite{transformer} to both generate new top-level samples and upsample them to the lower levels, accepting different features, such as lyrics, as conditioning. While autoregressive systems can achieve satisfying audio quality and long-range coherence, they suffer from extremely slow generation, as audio samples are produced sequentially. For example, Jukebox requires more than eight hours to generate one minute of audio on a V100 GPU. As an exception, RAVE \cite{rave} achieves real-time synthesis by encoding raw audio of a specific domain with a variational autoencoder \cite{vae} into a compact latent space and using a lightweight autoregressive model to generate codes: however, the short receptive field of the autoregressive model does not allow to model dependencies over distant time windows in the generated audio.

Non-autoregressive models avoid the slow sequential generation, but are mostly employed for \emph{conditional} audio synthesis. For example, several works focus on the task of inverting a low-dimensional audio representation (often a mel-spectrogram) back to the original waveform, which constitutes a building block of modern text-to-speech (TTS) systems \cite{melgan,hifigan,tacotron2}. 
In contrast, literature on long-form non-autoregressive unconditional audio generation is scarce. Systems such as WaveGAN and SpecGAN \cite{wavegan}, GANSynth \cite{gansynth}, DrumGAN \cite{drumgan} and MP3Net \cite{mp3net} attempt to generate audio of a fixed length using various architectures of Generative Adversarial Networks \cite{gan} (GANs). WaveGAN and SpecGAN represent the first works in which a GAN is successfully applied to audio, in the waveform and spectrogram representations, respectively. GANSynth generates instantaneous frequency (IF) and magnitude of spectrograms with high frequency resolution of short monophonic instrument notes \cite{nsynth}, showing that generating IFs and magnitudes instead of waveforms is advantageous for highly harmonic sounds. DrumGAN synthesizes drum sounds using the real and imaginary components of a complex STFT spectrogram and demonstrates the effectiveness of this audio representation, first introduced in \cite{comparing}. Finally, MP3Net achieves minute-long coherent piano music generation using Modified Discrete Cosine Transform (MDCT) spectrograms as audio representations and Progressive GAN \cite{progan} as the model: however, the generated samples suffer from low perceived audio quality. To the best of our knowledge, UNAGAN\cite{unagan} is the only non-autoregressive GAN-based system capable of generating audio of arbitrary length. The model takes a sequence of noise vectors as input and uses a hierarchical structure to achieve short-term coherence in the generated mel-spectrograms, which are then inverted to waveform using a pretrained MelGAN vocoder\cite{melgan}. However, the model only generates single-channel audio and the independent sampling of noise vectors cause the generated samples to lack coherence through time.

Contrary to autoregressive models, the majority of GAN-based unconditional audio generation models are only able to synthesize audio samples of a fixed length.
However, in the field of computer vision, several recent works propose models capable of generating images of arbitrary size, by synthesizing single image patches in parallel and assembling them into the final image. This process results in a fast and efficient generation of images on modern hardware. The two most notable contributions to this line of work are InfinityGAN \cite{infinitygan} and ALIS \cite{alis}. InfinityGAN generates in parallel single patches that are coherent with each other by disentangling global appearance, local structures and textures, which are then fed into a generator, together with coordinate information, to synthesize the final patch. ALIS proposes the use of latent vectors as anchor points for the coordinate system of the model: the resulting generator is equivariant and can thus produce coherent patches from interpolations of different latent codes. However, both of the methods rely on prior knowledge regarding the particular image domain that is being generated: the experiments are conducted on a dataset of images of landscapes, where the image features of a single patch are spatially invariant on the horizontal dimension and thus permit infinite length generation along the horizontal axis.

The process of generating sequences of encoded representations has been explored in different previous works \cite{vqvae,vqvae2,vqgan} for both image and audio data. However, these works focus on encoding samples to a discrete set of codes using vector-quantized variational autoencoders, and propose to model sequences of codes using autoregressive models. On the other hand, \cite{molecule} proposes to autoencode molecules with a basic autoencoder, to then generate sequences of continuous-valued latent vectors with a GAN. This approach manages to circumvent the problematic behaviour of GANs when applied on discrete data \cite{gumbel}, in this case molecules in the SMILES format. Similarly to this work, we propose to generate latent representations of audio with the aim of making the generation and training process faster, and achieving coherent generated samples over long time windows.


\section{Method}\label{sec:method}
Let $\mathbf{x}=\{\mathit{x_1},...,\mathit{x_T}\}$ be the waveform of an audio sample. We aim to encode a waveform $\mathbf{x}$ into a sequence of latent vectors $\mathbf{c}=\{\mathit{c_1},...,\mathit{c_{T/r_{time}}}\}$ with time compression ratio $r_{time}$, sampled at a lower sampling rate than the original waveform. We use an autoencoder model to perform this task, such that a reconstruction of the original waveform can be obtained from the encoded latent vectors.

We then aim to model the distribution $\mathit{p(\mathbf{c})}$ with a Generative Adversarial Network (GAN). We employ a latent coordinate system that is used as conditioning for the generator $\mathit{G}$ to generate sequences of latent vectors of arbitrary length. We additionally condition the generator with a variety of conditioning signals, such that the generation process can be guided by human input. Finally, the generated sequence of latent vectors is inverted to a waveform with the previously trained decoder.

\begin{figure*}[t]
\centering
\includegraphics[width=\textwidth]{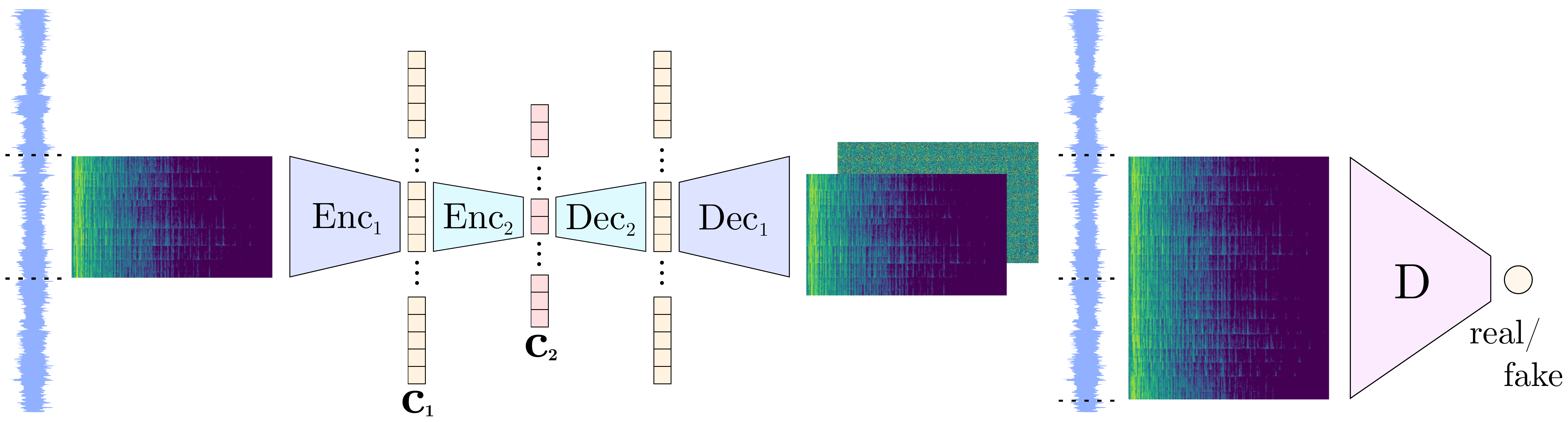}
\caption{The proposed 2-level audio autoencoder. A log-magnitude spectrogram is used as the encoder input, while the decoder outputs magnitude and phase spectrograms which are then inverted with iSTFT to the waveform domain. A discriminator evaluates the magnitude spectrogram of two adjacent excerpts passed through the autoencoder. This training process removes both phase errors (which would manifest after the iSTFT and STFT) and boundary artifacts.}
\label{fig:model}
\end{figure*}

\subsection{Audio Autoencoder}\label{subsec:autoencoder}
Considering the inherent high dimensionality of waveforms, generating long sequences of raw audio samples is prohibitively expensive.
A frequently used audio representation in the field of speech processing and music information retrieval is the Short-Time Fourier Transform (STFT) spectrogram: while the phase component of the spectrogram is usually discarded, in case of audio synthesis applications both magnitude and phase components are necessary to perform the inverse STFT (iSTFT) and obtain a waveform.

We design an audio autoencoder with the aim of minimizing inference and training time while maximizing the compression ratio allowing to reconstruct samples with reasonable accuracy. Our proposed autoencoder takes a log-magnitude STFT spectrogram as input, and outputs magnitude and phase spectrograms which can be inverted to a waveform. Parallel to our work, iSTFTNet \cite{istftnet} also proposes to improve the inference speed of the model by generating magnitude and phase of a STFT spectrogram: however, they only report experiments using spectrograms with very high temporal resolution and low frequency resolution, while our proposed autoencoder reconstructs spectrograms with low temporal resolution and high frequency resolution. This should result in an even higher inference speed for similarly-sized models. In practice, we separately train two stacked autoencoders; this allows a higher compression ratio with satisfactory reconstruction quality, especially for more complex music domains. Similarly to RAVE \cite{rave}, we utilize a two-step training process:

\subsubsection{First training phase}\label{subsec:1stphase}
We first train the model to autoencode log-magnitude spectrograms, not producing phases for now. We use a L1 loss function for the reconstruction task:
\begin{equation*}
\mathcal{L}_{(Enc,Dec),rec} = \mathbb{E}_{s \sim p(s)}||Dec(Enc(s))-s||_1
\end{equation*}
where $Enc$ and $Dec$ are the encoder and decoder, and $s$ is a log-magnitude spectrogram of a waveform $w$.

\newcommand{\origspect}{s}
\newcommand{\recospect}{\tilde{s}}
\newcommand{\origwav}{w}
\newcommand{\recowav}{\tilde{w}}

\subsubsection{Second training phase}\label{subsec:2ndphase}
In the second phase, we freeze the encoder weights and have the decoder produce a phase spectrogram as well, such that we can reconstruct a waveform through an iSTFT. We add an adversarial objective to aid the modeling of both the magnitudes and phases, ensuring the waveform is of perceptually satisfactory quality.
Since directly modeling phase spectrograms with deep learning models is known to be difficult \cite{gansynth,comparing}, we propose to model the phases indirectly, by encouraging waveforms whose magnitude spectrogram must appear realistic.
Specifically, we compute a log-magnitude spectrogram $\recospect$ from the reconstructed waveform $\recowav$: \begin{align*}
\recowav&=iSTFT(Dec(Enc(\origspect)))\\
\recospect&=\log(|STFT(\recowav)|^2+ \epsilon )
\end{align*}
The reconstructions $\recospect$ are fed to a discriminator $D$, using the hinge loss \cite{geometricgan} to distinguish them from originals $\origspect$:
\begin{align*}
\mathcal{L}_D=&-\mathbb{E}_{\origspect \sim p(\origspect)}[min(0,-1+D(\origspect))] \\ &-\mathbb{E}_{\origspect \sim p(\origspect)}[min(0,-1-D(\recospect))]
\end{align*}
The decoder is trained to fool the discriminator:
\begin{align*}
\mathcal{L}_{Dec,adv}=&-\mathbb{E}_{\origspect \sim p(\origspect)}D(\recospect)
\end{align*}
Note that we can calculate spectrograms from the reconstructed waveforms with different hop size and window length than used for the spectrograms fed to the autoencoder. We leverage this by including the multi-scale spectral distance \cite{ddsp,rave} in the objective of the decoder:
\begin{multline*}
\mathcal{L}_{Dec,ms}=\mathbb{E}_{\origwav \sim p(\origwav)}\sum_{hop}^N \log(|| \, |STFT_{hop}(\origwav)| \\ -|STFT_{hop}(\recowav)| \, ||_1)
\end{multline*}
where $hop$ indicates a choice of $hop\_size$ and $fft\_size$.

In total, we train the discriminator with $\mathcal{L}_D$, and the decoder with a linear combination of three losses:
\begin{align*}
\mathcal{L}_{Dec}&=\mathcal{L}_{Dec,adv} + \lambda_{rec} \mathcal{L}_{Dec,rec} + \lambda_{ms} \mathcal{L}_{Dec,ms}
\end{align*}

\begin{figure*}[t]
\centering
\includegraphics[width=0.85\textwidth]{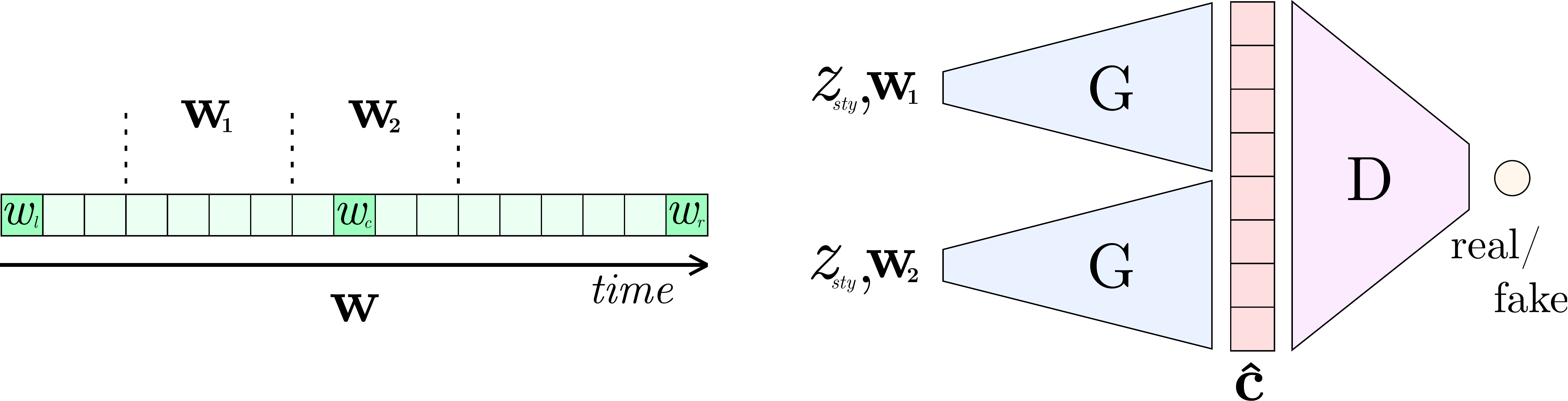}
\caption{The proposed latent GAN training process. Two adjacent latent coordinate sequences are randomly cropped from the linear interpolation between three anchor vectors. They are then used as input to the generator, together with a shared style vector and a conditioning signal in the case of a conditional model. The discriminator takes as input the two concatenated generated sequences and a real sequence of latent vectors. }
\label{fig:gan}
\end{figure*}

\subsection{Latent Coordinate System}\label{subsec:coordinate}
We use a GAN to model sequences of latent vectors produced by a trained audio encoder. In order to generate independent audio samples that can be seamlessly concatenated with each other along the temporal axis, we condition the generator with the latent coordinate system proposed by \cite{alis}, originally introduced to generate landscape images of infinite width. Specifically, during training we sample three noise vectors $\mathit{w_l},\mathit{w_c},\mathit{w_r}$ with dimension $d$ that are used as anchor points (left, center, right anchors) to guide the generation process. With $seq\_len$ being the length of the sequence of latent vectors that is produced by the generator, we linearly interpolate the three anchor vectors to create a sequence of coordinate vectors of length equal to $4 \cdot seq\_len +1$:
\begin{equation*}
    \mathbf{w}=[\mathit{w_l},...,(1-k)\mathit{w_l}+k\mathit{w_c},...,\mathit{w_c},...,\mathit{w_r}] \in \mathbb{R}^{4seq\_len +1 \times d}
\end{equation*}
To generate sequences that are temporally coherent with each other, we follow \cite{alis}: we randomly crop a sequence $\mathbf{w_{12}}$ of $2 \cdot seq\_len$ coordinate vectors from $\mathbf{w}$, divide it into two sequences $\mathbf{w_1},\mathbf{w_2}$ with length $seq\_len$, generate two patches using each sequence as conditioning, concatenate the two patches along the time axis, and feed the resulting generated sample of length $2 \cdot seq\_len$ to the discriminator. This process is illustrated in Figure \ref{fig:gan}. It allows the generator to align the sequence of latent coordinates with the generated sequence of latent vectors. Specifically, the discriminator forces the generator to learn that adjacent sequences of latent coordinates must result in adjacent sequences of latent vectors, which can be temporally concatenated resulting in a coherent final sample without artifacts at the boundaries of the generated patches.

Similarly to InfinityGAN \cite{infinitygan}, when generating adjacent sequences of latent vectors, we also condition both generations on a single random vector $\mathit{z_{sty}}$: during the learning process, this vector serves as conditioning for the global style of the generated samples. Specifically, while the latent coordinate vectors allow the generator to produce sequences of latent vectors that can be seamlessly concatenated along the temporal axis, the global style vector allows the final concatenated sequence of possibly infinite length to be stylistically coherent throughout. Without the global style vector, any temporal context available to the generator would completely change every $4 \cdot seq\_len$ samples of a sequence, resulting in a final generated sample which continuously changes style through time.

Formalized, we have
\begin{equation*}
    \mathbf{\hat{c}}=concat[G(\mathbf{w_1},\mathit{z_{sty}}),G(\mathbf{w_2},\mathit{z_{sty}})],
\end{equation*}
where $\mathbf{\hat{c}}$ is a stylistically and temporally coherent sequence of latent vectors of length $2 \cdot seq\_len$, and $G$ is the generator model.

At inference time, a latent coordinate sequence of the desired length is created. The coordinate sequence is prepared in the same way as during the training phase, by placing a latent anchor vector at positions that are multiples of $2 \cdot seq\_len$, and by linearly interpolating these anchor vectors to calculate in-between vectors. A single random global style vector is also sampled. Each generation considers a $seq\_len$ crop and the global style vector as conditioning, and finally all generated latent vectors are concatenated together in the appropriate order. This process can be performed in a parallel manner, thus resulting in a fast generation on modern hardware.


\begin{figure*}[t]
\centering
\includegraphics[width=.49\textwidth,trim=0 201 0 0,clip]{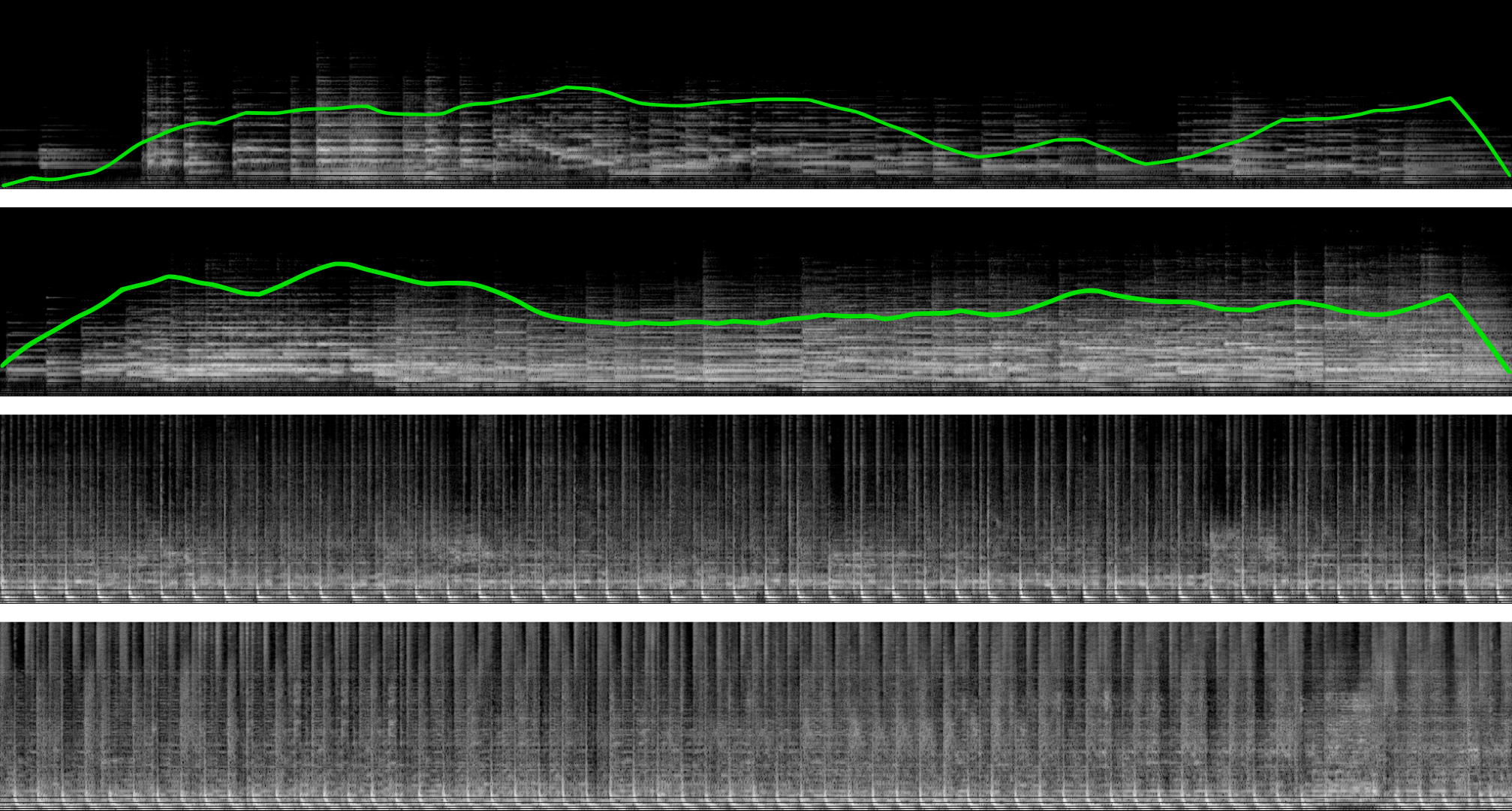}
\includegraphics[width=.49\textwidth,trim=0 133 0 68,clip]{figs/piano_techno_cond.png}
\includegraphics[width=.49\textwidth,trim=0 65 0 136,clip]{figs/piano_techno_cond.png}
\includegraphics[width=.49\textwidth,trim=0 -3 0 204,clip]{figs/piano_techno_cond.png}
\caption{Log-melspectrograms of generated piano and techno samples from the conditional models. For the piano samples (top row), we indicate the corresponding note density conditioning with a green line. Note density signals were generated using a random walk algorithm. The tempo used as conditioning for the techno samples (bottom row) is 120 bpm and 160 bpm, respectively. Each sample is 23 seconds long. Visit \href{https://marcoppasini.github.io/musika}{marcoppasini.github.io/musika} to listen to the examples.}
\label{fig:cond}
\end{figure*}

\section{Implementation Details}\label{sec:implementation}

\subsection{Audio Autoencoder Architecture}\label{subsec:aearch}
We first train an audio autoencoder with a relatively low compression ratio, then train a second-level autoencoder that encodes the first-level latent vectors, as shown in Figure \ref{fig:model}. During the training of the second-level autoencoder, we utilize the same training strategy and objective as explained in section \ref{subsec:autoencoder}, by propagating gradients through the frozen weights of the previously trained first-level decoder and adversarially discriminating between samples reconstructed by both decoders and samples reconstructed by only the first-level decoder. Both model architectures are fully convolutional, and we do not use any padding in both encoders, such that possible boundary artifacts in the encoded representations are avoided. We utilize 1d-convolutions considering the frequency bins as different channels for both encoders and decoders: this is usually not efficient regarding total number of model parameters when compared to using 2d-convolutions across the two spectrogram dimensions, but can result in a much faster inference time. We use 2d-convolutions for the discriminator, as inference time for this model is not a priority. We use $tanh$ as the activation for the bottleneck layer of both encoders. Regarding the multi-scale spectral distance loss, we use $hop\_size \in [64,128,256,512]$ and we always choose $fft\_size = 4 \cdot hop\_size$, while the discriminator takes as input log-magnitude spectrograms calculated with $hop\_size=256$ and $fft\_size=6 \cdot hop\_size$. As proposed by \cite{cocogan,melganvc}, two consecutive reconstructed spectrograms are concatenated along the temporal dimension and fed to the discriminator, such that concatenated reconstructions do not suffer from boundary artifacts. During training, spectrograms calculated from 0.76\,s of audio are used as input to both autoencoders. We use spectral normalization \cite{spectral} on the weights of the discriminator. Regarding the training loss weights, we use $\lambda_{rec}=1$ and $\lambda_{ms}=4$. We choose Adam \cite{adam} as the optimizer with learning rate of $0.0001$ and $\beta_1=0.5$, and train the first-level autoencoder for 1 million iterations with batch size of 32 for both training phases, and the second-level autoencoder for 400k iterations with batch size of 32 for both training phases.

\subsection{Latent GAN Architecture}\label{subsec:architecture}
We choose to adapt the FastGAN \cite{fastgan} architecture to our specific task. The FastGAN architecture promises fast convergence with limited amounts of data. To achieve this, it proposes a Skip-Layer channel-wise Excitation (SLE) module in the generator, for more direct propagation of gradients, and proposes to strongly regularize the discriminator with an added self-supervised reconstruction objective. We adapt the proposed architectures to use 1d-convolutions instead of 2d-convolutions and we simplify the added reconstruction objective of the discriminator, by using a single lightweight decoder which reconstructs the whole input of the discriminator. Differently from FastGAN, we do not use Batch Normalization \cite{batchnorm} in both the generator and discriminator, while we apply the variation of Adaptive Instance Normalization \cite{adain} (AdaIN) called Spatially Aligned AdaIN (SA-AdaIN), originally proposed in \cite{alis}, after each convolutional layer in the generator. To generate stereo samples, the generator produces two latent vectors at each timestep, one for each audio channel, stacked on the channel axis. We use Cross Channel Mixing (CCM), first introduced in \cite{projgan}, to randomly mix channels of the stereo stacked latent vectors before being fed to the discriminator. In our experience, this technique helps reducing collapses during training. Both anchor and style vectors are sampled from a normal distribution with zero mean and unit variance, and have dimension $d$ of 64. We use R1 gradient penalty \cite{gp} as regularization, and Adam with learning rate of $0.0001$ and $\beta_1 = 0.5$ as the optimizer. We train for 1.5 mio iterations with a batch size of 32 for all experiments. Training takes 23\,h on a RTX 2080 Ti GPU.

\begin{table}[t]
\centering
 \begin{tabular}{lll}
  \toprule
  \textbf{Model (Faster than real-time)} & \textbf{GPU} & \textbf{CPU}\\
  \midrule
  Musika Uncond. Piano  & 972x & \textbf{40x} \\
  Musika Cond. Piano  & 921x & \textbf{40x} \\

  UNAGAN\cite{unagan} Piano  & 28x & 11x \\
  
  Musika Uncond. Techno  & \textbf{994x} & 39x \\
  Musika Cond. Techno  & 917x & 39x \\
  \bottomrule
 \end{tabular}
 \caption{Comparison of generation speed between the different models. For the Musika models, we include both the generation of the latent vectors and the decoding step to the waveform domain. We use a RTX 2080 Ti and a Ryzen 3950x as the GPU and CPU, respectively. We report the average of 100 trials.}
 \label{tab:speed}
\end{table}

\begin{table}[t]
\centering
 \begin{tabular}{ll}
  \toprule
  \textbf{Model} & \textbf{FAD}\\
  \midrule
  Musika Uncond. Piano  & \textbf{1.641} \\
  Musika Cond. Piano Rand.  & 2.150 \\
  Musika Cond. Piano Const. 0.15  & 2.584 \\
  Musika Cond. Piano Const. 0.30  & 3.400 \\
  Musika Cond. Piano Const. 0.45  & 4.389 \\
  Musika Cond. Piano Const. 0.60  & 4.839 \\
  Musika Cond. Piano Const. 0.75  & 5.434 \\
  UNAGAN\cite{unagan} Piano  & 11.183 \\
  \bottomrule
 \end{tabular}
 \caption{FAD evaluation for generated piano music. We evaluate conditional Musika models using different constant values of note density as conditioning. We notice that FAD increases with higher note density.}
 \label{tab:fad}
\end{table}

\section{Experiments}\label{sec:experiments}
Considering the relatively low compression ratio of the first autoencoder and thus its need to only encode low-level audio features, we find it possible to train a single universal model which we can later use for different music domains. As training data, we choose to use songs released and made freely available by South by SouthWest\footnote{\href{https://www.sxsw.com/festivals/music/}{https://www.sxsw.com/festivals/music/}} (SXSW) in occasion of their yearly conference. The current collection consists of 17k songs of various genres, and for this reason it represents a fitting choice for training our universal model. We use the LibriTTS corpus \cite{libritts} as additional training data, to steer the universal model into accurately synthesizing human voices, which are notoriously hard to model. Even though LibriTTS only contains speech, including it improves reconstructions of singing voice. We resample audio to 22.05\,kHz for all experiments. We use single channel audio to train the audio autoencoders, as the latent GAN is able to generate stereo samples by using latent representations of the two mono samples stacked in the channel dimension as training data. We use $r^1_{time}=256$ as the time compression ratio, which results in a sampling rate of the first-level latent representations of 190.22\,Hz. Each of the encoded latent vectors has a dimension of 128.


\subsection{Piano Music}\label{subsec:piano}
We use the MAESTRO dataset \cite{maestro}, consisting of 200 hours of piano performances, to train a second-level autoencoder and a latent GAN. The final time compression ratio achieved by both autoencoders is $r_{time}=4096$, which results in a sampling rate of the second-level latent representations of 11.89\,Hz. The dimension of each latent vector is 32. We train both an unconditional and a conditional latent GAN. For both models, the generator outputs latent vectors with $seq\_len=64$, which results in about 12 s of audio after decoding. For the conditional model, we apply the CNN-based onset detector \cite{onsetcnn} of the madmom Python library \cite{madmom} to all audio files in the dataset. We then use Gaussian Kernel Density Estimation (KDE) with bandwidth of 0.004 on the detected onsets to estimate a continuous note density signal for each sample. This signal is log-scaled between 0 and 1 and serves as a conditioning signal for the conditional (and thus controllable) GAN.

\subsection{Techno Music}\label{subsec:techno}
To evaluate the performance of the system on a more musically varied domain, we scrape 10,190 songs categorized with the ``techno'' genre from \textit{\href{https://jamendo.com}{jamendo.com}} and use them as training data. Considering the wide diversity of sounds that are present in the dataset, we train the second-level autoencoder with the same SXSW data used to train the first-level universal autoencoder. Comparing to what is achievable when training an autoencoder on a single and limited domain, such as piano music, a lower compression ratio is needed to reach a satisfactory reconstruction accuracy. However, this solution allows users to directly train a latent GAN on a new audio domain using the universal latent representations, without the need to train an autoencoder on the domain of interest. The final achieved time compression ratio is $r_{time}=2048$, which results in a sampling rate of the second-level latent representations of 23.78\,Hz. The dimension of each latent vector is 64. We train an unconditional and a conditional latent GAN model, both generating stereo latent vectors with $seq\_len=128$, resulting in about 12\,s of decoded audio. We use the Tempo-CNN framework\footnote{https://github.com/hendriks73/tempo-cnn} \cite{tempo-cnn} to estimate the global tempo of each song in the dataset. Tempo information is then used as conditioning for the conditional model.

\section{Results}\label{sec:results}
A comprehensive collection of generated audio samples is available on \href{https://marcoppasini.github.io/musika}{marcoppasini.github.io/musika}. Since current quantitative evaluation metrics are not able to assess the overall compositional and musical quality of generated music, we strongly encourage the reader to listen to the provided samples while reading the paper.

We report the generation speed of the system trained on the MAESTRO and on the techno datasets in Table \ref{tab:speed}, on both GPU and CPU. We also use the Frechét Audio Distance\cite{fad} (FAD) metric to quantitatively evaluate the quality of the generated piano samples in Table \ref{tab:fad}. A UNAGAN\cite{unagan} model that was trained on the same dataset is used as comparison. While our system is capable of generating stereo audio, UNAGAN can only produce single-channel audio. The unconditional model obtains the lowest FAD, while the conditional system results in higher FADs when using more intense note density values as conditioning. This is expected, since samples with low note density are more common than samples with high note density in the MAESTRO dataset. However, considering that audio is split in short 1\,s samples to calculate embeddings, FAD is not designed to evaluate overall musical and compositional quality of samples, and to the best of our knowledge there are no available quantitative metrics to evaluate these characteristics. Piano and techno samples generated by the system seem to often demonstrate long-range coherence and successfully keep a fixed general music style through time. Both conditional models successfully generate samples that are coherent with the conditioning signal, as can be seen in Figure \ref{fig:cond}.



\section{Conclusion}\label{sec:conclusion}
We proposed Musika, a non-autoregressive music generation system that generates raw-audio samples of arbitrary length much faster than real-time on a consumer CPU. An efficient hierarchical autoencoder allows to encode audio to a sequence of low-dimensional latent vectors, from which a waveform can be reconstructed. A GAN is then used to generate new sequences of latent vectors, using a latent coordinate system that allows for generation of samples of infinite length. A style conditioning vector is introduced to force the samples to be stylistically coherent through time. We successfully use the system to generate piano and techno music, and show that the generation process can be conditioned on note density and tempo information for piano and techno music, respectively. We finally show that the system achieves lower FAD than comparable systems on piano music generation while being faster.  
We release the source code and pretrained models, enabling users to generate samples of different music domains and test new conditioning signals with ease and using consumer hardware.
We see our system as solving an important technical challenge -- real-time music generation of sufficient quality, conditioned on user input -- and hope it can serve as a basis for interactive real-world applications and for research into human-AI co-creation.

\bibliography{ISMIRtemplate}

%
%
%
%
%

\end{document}